\documentclass[twoside]{article}
\usepackage[section]{algorithm}
\usepackage{amsthm,amsmath,amssymb,amsmath,amsfonts,graphicx,fancyhdr,algorithmic,longtable,listings,color}

\makeatletter
\newcounter{algorithmbis}
\renewcommand{\thealgorithmbis}{\thesection.\arabic{algorithmbis}}
\def\algorithmbis{\@ifnextchar[{\@algorithmbisa}{\@algorithmbisb}}
\def\@algorithmbisa[#1]{%
  \refstepcounter{algorithmbis}
  \trivlist
  \leftmargin\z@
  \itemindent\z@
  \labelsep\z@
  \item[\parbox{\textwidth}{%
    \hrule
    \hrule
    \noindent\strut\textbf{Algorithm \thealgorithmbis} #1
    \hrule
  }]\hfil\vskip0em%
}
\def\@algorithmbisb{\@algorithmbisa[]}

\makeatother

\def\enddemo{\qed \endtrivlist}
\expandafter\let\csname enddemo*\endcsname=\enddemo

\def\qedsymbol{\ifmmode\bgroup\else$\bgroup\aftergroup$\fi
  \vcenter{\hrule\hbox{\vrule
height.6em\kern.6em\vrule}\hrule}\egroup}
\def\qed{\ifmmode\else\unskip\nobreak\fi\quad\qedsymbol}

 \newtheorem{exm}{Example}[section]

  \newtheorem{ppro}{Procedure}[section]

\newcommand{\ra}[1]{\mathrm{rank}({#1})}  

 \font\ssr=cmss8 \font\sst=cmtt8 
 \font\ssi=cmti8

\title{\bf Computation of generalized inverses \\ using {$Php/MySql$} environment}
\begin{footnotesize}
\author{\frenchspacing
\bf  Milan B. Tasi\'c\footnote{Corresponding author\ }, Predrag S. Stanimirovi\' c, Selver H. Pepi\' c\\
{\ssi University of Ni\v{s}, Faculty of Science and Mathematics,}\\
{\ssi Vi\v segradska 33, 18000 Ni\v s, Serbia} \\
{\ssi E-mail:}\ \ {\sst milan12t@ptt.rs},\ {\sst pecko@pmf.ni.ac.rs}, \ {\sst p\_selver@yahoo.com}\\
}
\end{footnotesize}
\date{}
\begin{document}

\maketitle

\begin{abstract}
The main aim of this paper is to develop a client/server based model for computing the weighted Moore-Penrose inverse using the partitioning method as well as for storage of generated results.
The web application is developed in the $PHP$/$MySQL$ environment.
The source code is open and free for testing by using a Web browser.
Influence of different matrix representations and storage systems on the computational time is investigated.
The CPU time for searching the previously stored pseudo-inverses is compared with the CPU time spent for new computation of the
same inverses.

\frenchspacing \itemsep=-1pt
\begin{description}
\item[] AMS Subj. Class.: 15A09, 68P15, 68N15.
\item[] Key words: {\ssr PHP\/} script, {\ssr MySQL\/} database, Weighted Moore-Penrose inverse, Matrix storage formats.
\end{description}
\end{abstract}

\definecolor{listinggray}{gray}{0.9}

\section{Introduction} \setcounter{equation}{0}

Since the mid-1990s, there has been a surge of interest among the academics and practitioners in an open source software ($OSS$).
There are many successful projects in $OSS$ community, primarily the $Mozilla$ web browser, the $Linux$ operating
system, the $Apache$ web server, and to a lesser extent, $PHP$ \cite{Meloni} and the $Perl$ programming languages, as well as the $MySQL$ database.
$OSS$ has drawn the attention of users and developers because of its economic benefits \cite{Lee}.

\smallskip
$PHP$ is an open source software, and it is free for downloading and use.
Main characteristics of $PHP$ are described in \cite{Meloni}.

$MySQL$ has become the world's most popular open source database system because of its consistent fast performance,
high reliability and ease of usage.
Besides the fact that it is free, $MySQL$ offers a wide range of possibilities \cite{Greenspan,Williams}.

\smallskip
$SQL$ is the standard language used for querying and analysis of data in a relational $DBMS$ \cite{Elmasri}.
Unfortunately, $SQL$ has no vector and matrix computations.
Interesting code for matrix $SQL$ operations can be found in \cite{Robyn}.
The $SQL$ constructions and the $SQL$ primitives for data mining are proposed in \cite{Clear}.
However, these constructions do not offer adequate and flexible tools for matrix manipulation.
A graphical $SQL$ query generator and query operators with embedded matrix objects are disclosed in \cite{Egilsson}.

\smallskip
Implementation of some vector and matrix operations based on programming User-Defined Functions ($UDFs$) is studied in \cite{Ordonez}.
$UDFs$ represent a $C$ programming interface that allows the definition of scalar and aggregate functions that can be used in $SQL$.
$UDFs$ have several advantages and limitations. An $UDF$ allows fast evaluation of arithmetic expressions, memory manipulation,
using multidimensional arrays and exploiting all $C$ language control statements.

\smallskip
The organization of the paper is as follows.
Motivation of the paper is described in the second section.
In the third section, we describe application details.
In the fourth section, we develop data storage and computing system, which manipulate with different types of matrices, based on the $PHP/MySQL$ environment.
Functions and procedures are hardwired to a common matrix table.
Influence of different matrix representations in conjunction with storage systems to performances of the implemented software is considered through numerical experiments.
All test matrices and the results can be stored in files and the database on the server-side.
A few illustrative examples and comparative studies are presented in the last section.

\section{Preliminaries and motivation} \setcounter{equation}{0}

Some mathematical formulations and motivation are involved in this section.
Let $\Bbb C$ be the set of complex numbers, $\Bbb C^{m\times n}$
be the set of $m \times n $ complex matrices, and
${\Bbb C}^{m\times n}_r\!=\!\{X\in {\Bbb C}^{m \times n}\, :\,\,\, \ra{X}\!=\!r\}$.
For any matrix $A\in\Bbb {C}^{m \times n}$ and positive definite
matrices $M$ and $N$ of the orders $m$ and $n$ respectively, consider the following
equations in $X$, where $*$ denotes conjugate and transpose:
$$\begin{array}{ll}
  (1)\qquad \   AXA=A & (2)\qquad XAX\! =\! X \\
  (3M)\quad  (MAX)^*=MAX & (4N)\quad (NXA)^*\! =\! NXA.
\end{array}$$

The matrix $X$ satisfying (1), (2), (3M) and (4N) is called the weighted Moore-Penrose inverse of  $A$, and it is denoted by $X=A_{MN}^{\dagger}$.
In the particular case $M=I_m$ and $N=I_n$, the matrix $X=A_{MN}^{\dagger}$ comes to
the Moore-Penrose inverse of $A$, and it is denoted by $X=A^{\dagger}$.

\smallskip
The Greville's {\it partitioning method\/} for numerical computation of generalized inverses
is introduced in \cite{Gre}.
The following computational experience is delivered in \cite{Layton}:
"when applied to a square, fully populated, non-symmetric case, with independent
columns, the Greville's algorithm found that the approach can be up to eight times faster than the conventional approach of using the SVD; rectangular cases are shown to yield similar levels of a speed increase".
Due to its computational dominance, this method has been extensively applied in many mathematical areas, such as statistical inference, filtering theory, linear estimation theory, optimization and more recently analytical dynamics \cite{Udwadia3N} (see also \cite{Kurmayya}).
An application of the partitioning method in a direct approach for computing the gradient of the pseudo-inverse is presented in \cite{Layton}.
It has also found wide applications in the database and the neural network computation \cite{Mohi}.
In the paper \cite{Itiki}, the sequential determination of the Moore-Penrose generalized inverse matrix by dynamic programming is applied to the diagnostic classification of electromyography signals.

\smallskip
The Greville's partitioning algorithm is extended in many papers.
Wang in \cite{Wang} generalizes Greville's method to the weighted Moore-Penrose inverse.
The algorithm for computing the Moore-Penrose inverse of one-variable polynomial and/or rational matrices,
based on the Greville's partitioning algorithm, is established in \cite{Stanimirovic2}.
An extension of the results from \cite{Stanimirovic2} to the set of two-variable rational and polynomial
matrices is introduced in the paper \cite{Stanimirovic3}.
In the paper \cite{Tasic1} we extended the Wang's partitioning method from \cite{Wang} to the set of
one-variable rational and polynomial matrices. An efficient implementation of the algorithm
introduced in \cite{Tasic1} is established in \cite{Petkovic}.
An implementation of the algorithm introduced in \cite{Tasic2} for computing generalized inverses is based on the $LU$ factorization of the matrix product.

\smallskip
For the sake of completeness, we restate the algorithm introduced in \cite{Wang}, applicable to rational and constant matrices.
The algorithm is quite appropriate for computation of the weighted Moore-Penrose inverse as well as the Moore-Penrose inverse and regular inverse.

\begin{algorithmbis}[(G.R. Wang, Y.L. Chen.)\ Computing the weighted M-P inverse $A_{M,N}^\dag$.]\label{algwang}
\begin{algorithmic}[1]
\REQUIRE Let $A\in \Bbb{R}^{m\times n}$, $M$ and $N$ be p.d. matrices of the order $m$ and $n$ respectively.
\STATE $A_1=a_1.$
\IF {$a_1=0,$}
\STATE $X_1=(a_1^* M a_1)^{-1} a_1^* M;$
\ELSE
\STATE $X_1=0.$
\ENDIF
\FOR{$k=2$ to $n$}
\STATE $d_{k} \!=\!X_{k-1}a_k,$ $c_k=a_k-A_{k-1}d_k,$
\IF {$c_k\ne0,$}
\STATE $b_k^*=(c_k^* M c_k)^{-1} c_k^* M, \ {\bf goto} \ Step\ \ref{sest},$
\ELSE
\STATE $\delta_k=n_{kk}+d_k^* N_{k-1} d_k - (d_k^* l_k+l_k^* d_k)-l_k^*(I-X_{k-1}A_{k-1})N_{k-1}^{-1} l_k,$
\STATE $b_k^*=\delta_k^{-1}(d_k^*N_{k-1} -l_k^*) X_{k-1},$
\ENDIF
\STATE $X_k=\left[\begin{array}{cc} X_{k-1}-(d_k+(I-X_{k-1}A_{k-1})N_{k-1}^{-1} l_k)b_k^*\\
b_k^*\end{array}\right].\label{sest}$
\ENDFOR
\STATE {\bf return} $A_{MN}^\dag=X_n.$
\end{algorithmic}
\end{algorithmbis}

Comparison of searching in large databases with recomputation in recursive algorithms is one of the questions on which we will focus in this article.
The partitioning method is chosen in our article because of its recursive structure.
Substantial effort is dedicated to a tight coupling of database and matrix computations, particularly in the inversion problem.
Our idea is to memorize all input and resulting matrices in database tables, and as we need some of them, simply find results in appropriate table.
Obviously, it is more efficient to find results on the server rather than perform recomputation, in the case when input matrices and requested operations are stored on the server side. This approach skips the computational logic and reduces the CPU time.
We discuss the implementation of Algorithm \ref{algwang} on a numerical DBMS and present experimental results
demonstrating the performance benefit.

\smallskip
Many scientists ignore the fact that most of data that are subject to analysis has been already
exploited and can be stored in database systems, as well as that database systems provide powerful mechanisms for accessing, filtering and indexing data.
We develop an application which memorizes all input matrices and results of performed matrix computations.
The main challenge is at first to identify appropriate matrix representations and secondly to implement them using numerical DBMS extension facilities, e.g. storing, indexing, searching, etc.
To achieve our aim, we implemented and tested different methodologies for storing matrices in the database and systems of the matrix representation (application logic) in the three-tier architecture.

\smallskip
We develop an algorithm for storing every input matrix and results of computations by a database system.
The algorithm is applied in order to develop the client/server Web application for computing the Moore-Penrose inverse, the weighted Moore-Penrose inverse as well as the fundamental matrix operations.
Based on the analysis of available algorithms and our previous work in this area
(see \cite{Petkovic,Stanimirovic3,Stanimirovic2,AProblem,Tasic1,Tasic2,Tasic3,Tasic4}),
we have identified operations, which are useful for a number of similar algorithms.

\begin{figure}[ht]
  \begin{center}
    \includegraphics[width=7.2truecm]{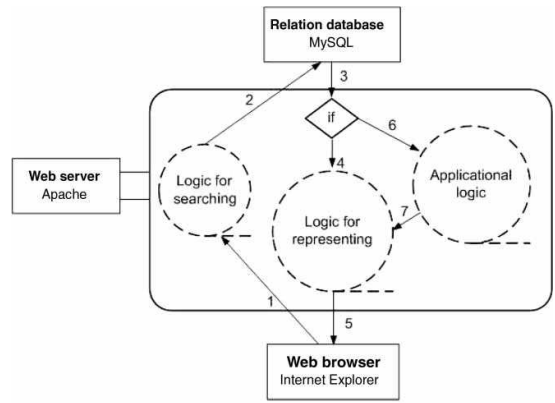}
    \caption{A computing machine on the middle tier} \label{FFig1}
  \end{center}
\end{figure}

\smallskip
The following figure illustrates our idea.
We compare the input matrices (entered in the web forms) and required matrix operations with matrices and operations in the database table.
If these matrices and required operations exist in a database table, the computational logic is determined by the steps $(1,2,3,4,5)$.
Otherwise, the computational logic follows the path $(1,2,3,6,7,5)$, as it is illustrated on Figure \ref{FFig1}.

\vfill\eject

An extension of $DBMS$ with fundamental vector and matrix operators supported by programming in the $PHP/MySQL$ environment is studied.
We experimentally compare the influence of different matrix representations and storage systems in computation of the weighted Moore-Penrose inverse using client-server architecture and $SQL$, with respect to performance, ease of use, flexibility and scalability.
The $PHP$ scripting language is used as our middle-tier scripting language in the three-tier architecture model of the web database application.

\smallskip
The research questions, we should answer in this article are the following:

- Can $PHP/MySQL$ environments help in writing common matrix operations aimed for computing the generalized inverse?

- Can we take advantages of the $PHP$ language to implement vector and matrix operations in a database?

- How different matrix representations, implemented using client-server architecture and $SQL$,
   enhances to performance, ease of use, flexibility and scalability of algorithms for computing ge\-ne\-ra\-li\-zed inverses?

\smallskip
In the present article we, continue the idea used in \cite{AProblem}, where the matrix database storage is used in the pseudo-inverse computation.
Besides the routines which compute the weighted Moore-Penrose inverse, we develop a set
of routines for implementation of the matrix library.
In this way, we also continue results from \cite{Karaman}, where a dynamic parallel matrix library is introduced.
Our fundamental matrix operations include addition, subtraction, multiplication of two matrices, computation of determinants, as well as the pseudo-inverse and the weighted pseudo-inverse computations.
All these operations are considered for both dense and sparse matrices, with a possibility of expansion on diagonal,
triangular and symmetric sparse matrices.
We also overcome results from the paper \cite{Nash}, where a set of {\ssr FORTRAN} subroutines for testing computer programs aimed for computing the generalized inverse is presented.
Usage of the programming language $PHP$ overcomes bounds of the {\ssr FORTRAN} and {\ssr C} subroutines from \cite{Nash} and \cite{Karaman} in the following possibilities:

1. Usage of the Web oriented programming paradigm available in $PHP$ in conjunction with $HTML$ and $XML$;

2. Usage of the object oriented programming ($OOP$);

3. Possibility to use the data storage system available in $MySQL$ database system;

4. Usage of the Apache Web server.

\smallskip
Therefore, the main idea is to provide an appropriate client-server application, in the free open source
$PHP/MySQL$ development environment, utilizing the minimum of resources: Internet browser and the operating system.

\section{Application details}\setcounter{equation}{0}

Database developer $SQL$ and $SQL$ Server does not support direct operations on matrices.
However, because tables and matrices share the same structures, $SQL$ allows easy manipulations with matrices.
The present article demonstrates a few $SQL$ techniques for performing basic matrix operations.
As we have mentioned, our idea is to implement a client/server application supported by the database usage,
in order to find a solution which is already placed in the database.
Our application makes possible operations with one, two and/or three arguments.
The application is a combination of $PHP/MySql$ elements and implements
special database operations, which support the $SQL$ software implementation of a wide range of algorithms.

\begin{figure}[ht]
  \begin{center}
    \includegraphics[width=7.4truecm]{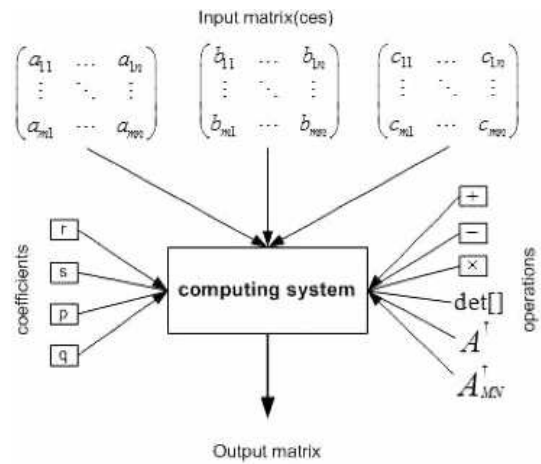}
    \caption{The matrix computation system} \label{FFig2}
  \end{center}
\end{figure}

Figure \ref{FFig2}. shows how the application works.

Each input matrix is, firstly, transformed into a string, with the elements separated by commas.
Then the generated strings, subject to the required operations, are compared with data stored in the database.
If the searching criteria are successfully ended, then a recomputation is not needed, and result will be displayed immediately in an appropriate web form.

The possibilities offered by the application are the following:

- \textbf{unary operations}: unary matrix operations, where only a single matrix and an optional coefficient is used to produce a unique result.
($A^{-1}$, $rA$, $A^{\dagger}$,...);

- \textbf{binary operations}: binary matrix operations, involving two matrices ($A+B$, $A^{p}\times
B^{q}$,...);

- \textbf{ternary operations}: operations involving three matrices, such as the computation of the weighted Moore-Penrose inverse $A^{\dagger}_{MN}$.

\smallskip
User can define input matrices in two ways:

- by element-wise entering values for elements of the matrices;

- by uploading $txt$ file, which contains matrix entries, under the assumption that matrix rows match with lines of the file and matrix elements are separated with the blank character.

\begin{figure}[ht]
  \begin{center}
    \includegraphics[width=14.6truecm]{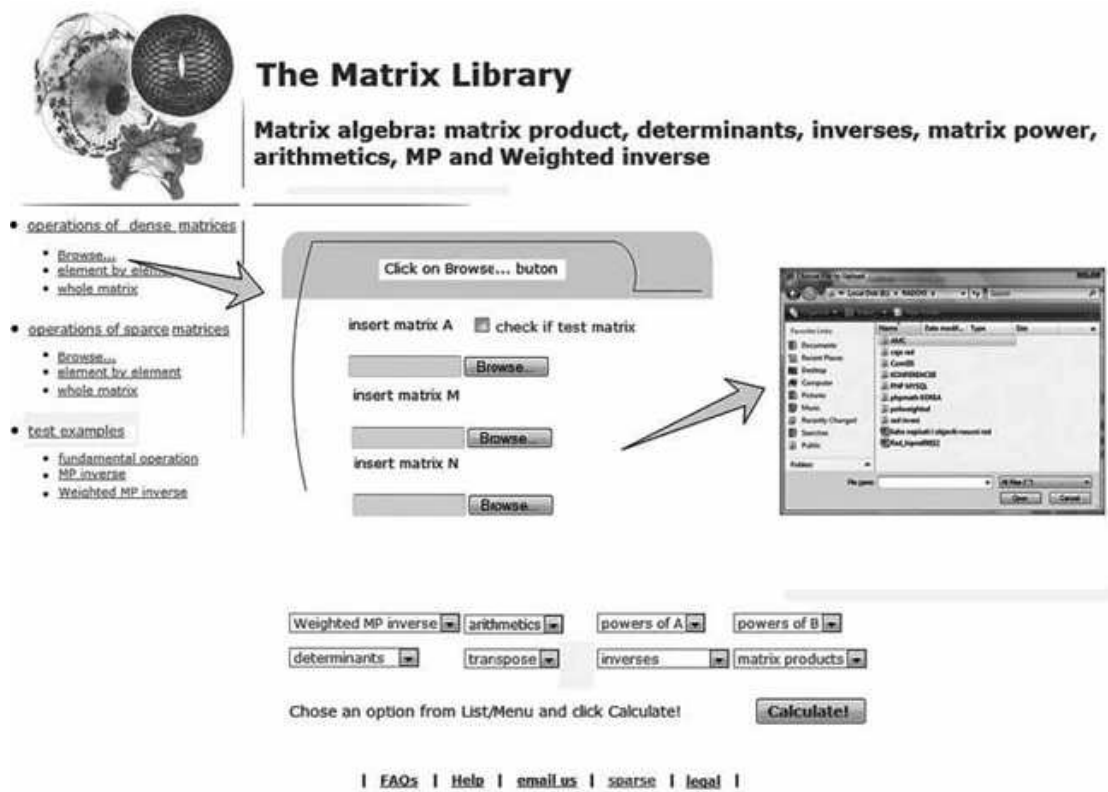}
    \caption{Web interface of the application} \label{fhaos3}
  \end{center}
\end{figure}

For this approach has been developed suitable Web interface illustrated in Figure \ref{fhaos3}.

\vfill\eject

How does the application work? General algorithm is defined globally based on the next three steps.

\smallskip
\textit{The first step}. The user chooses a matrix operation that will be implemented, via web browsers.

\begin{figure}[ht]
  \begin{center}
    \includegraphics[width=10.8truecm]{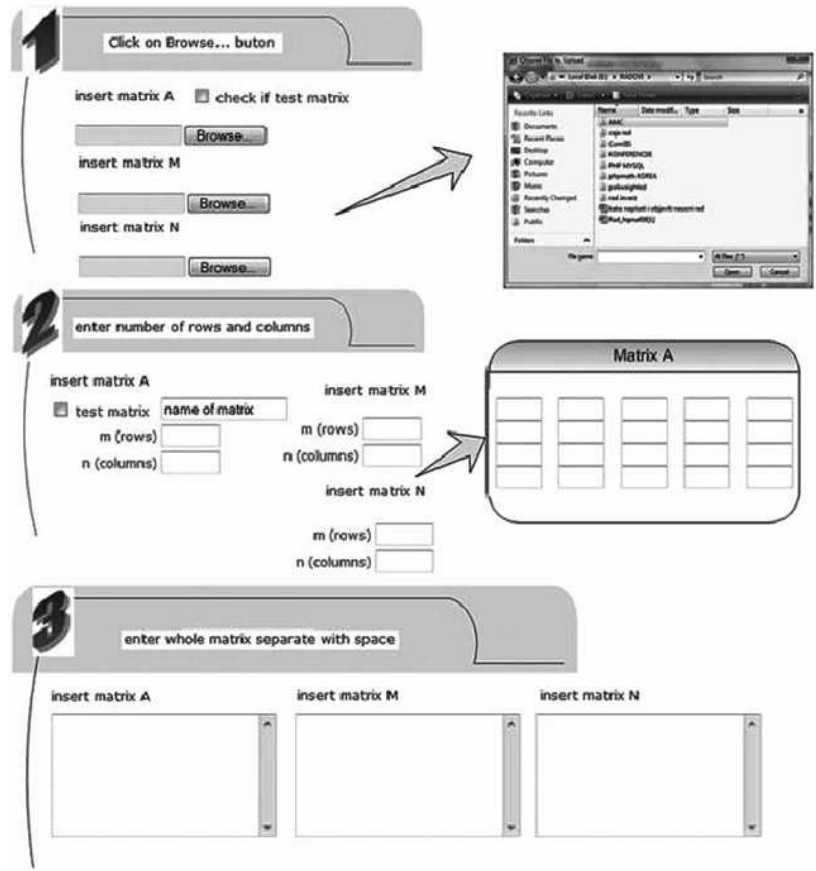}
    \caption{An example how to input matrix elements} \label{FFig4}
  \end{center}
\end{figure}

\smallskip
\textit{The second step}. Select the type of matrix that will be proceeded and define entries.
There are three types of matrices for processing: dense, sparse or test matrices.
More, it is possible to define entries of the input matrix in one of the next three ways, as it is illustrated on Figure \ref{FFig4}.

- loading the matrix from an existing textual file, using the button $Browse$.
Matrix is stored in memory as $txt$ file, assuming that the elements of the matrix are separated with
a comma, and the matrix rows are separated by the new line.

- entering the number of rows, columns and later values of the elements.

- entering a complete matrix in the text area box, assuming that its entries are separated by the blank character.

\smallskip
\textit{The third step}: Generate and preview the solution.

\vfill\eject

All the once handled matrices, chosen operations and the generated results of processing are stored on the server in the appropriate form in \textit{txt} files.
In this way, it is possible to avoid recomputations and download solutions of previously performed operations.
Users are also allowed to use ready-made test samples.

\section{The Storage system}\setcounter{equation}{0}

The matrices and their storage representations implemented in the paper can be classified in the following two groups.

\smallskip
{\bf Randomly generated and test matrices} are stored in the database in two different storage systems.
The row matrix format ($R$ format) assumes that all matrix elements are placed in a vector, so that
the matrix is represented as a string containing values of the matrix elements separated by the comma.
On the other hand, the $mR$ format represents a matrix in a database table under the assumption that the number of records in the table is equal to the number of rows of the matrix.

\smallskip
{\bf Sparse unstructured matrices} are matrices whose entries are equal to zero
for the most part and possess a distribution of the nonzero entries which do not match any specific pattern.
There are various storage schemes, which minimize the memory space and computational requirements, by storing and
performing arithmetic with only the nonzero elements.
The simplest sparse matrix storage structure is the {\it Coordinate Format ($COO$ format)} \cite{Chong},
where the matrix is stored in three appropriate vectors, which represent the underlying sparse structure.
The first vector (resp. the second vector) stores the row indices (resp. the column indices) for all non-zero entries.
The third vector stores non-zero entries of the sparse matrix.

\smallskip
The $SQL$ codes for creating the data structure and tables on $MySQL$ database server are described as follows.
The {\it CREATE TABLE} statement has three parts:

- A table name succeeds the $CREATE$ $TABLE$ statement.

- A list containing attribute names, types, and modifiers succeeds are placed between the parenthesis.

- A key list follows after the attribute list between the parenthesis.

In our implementation, we use only two different names for tables, namely $matrices\_in$ and $matrices\_out$;

\begin{footnotesize}
\begin{lstlisting}
# Table structure for table `matrices_in`
CREATE TABLE matrices_in (
  id_in int(11) NOT NULL auto_increment,
  elements_in longtext NOT NULL,
  dimension tinytext NOT NULL,
  test tinytext NOT NULL,
  sparse tinytext NOT NULL,
  PRIMARY KEY  (id_in)
) TYPE=MyISAM;

# Table structure for table `matrices_out`
CREATE TABLE matrices_out (
  id_out int(11) NOT NULL auto_increment,
  elements_out longtext NOT NULL,
  operation tinytext NOT NULL,
  matrix_I int(11) NOT NULL default '0',
  matrix_II int(11) NOT NULL default '0',
  matrix_III int(11) NOT NULL default '0',
  r tinyint(4) NOT NULL default '0',
  s tinyint(4) NOT NULL default '0',
  p tinyint(4) NOT NULL default '0',
  q tinyint(4) NOT NULL default '0',
  PRIMARY KEY  (id_out)
) TYPE=MyISAM;
\end{lstlisting}
\end{footnotesize}

The fields used in the database tables are described as follows.

- $id\_in$ or $id\_out$: identification number, defined as $auto\_increment$;

- $elements\_in$ or $elements\_out$: string of type $longtext$, containing matrix entries separated with the comma character.

- $dimension$: string of the form '$m$ x $n$' which contains dimensions $m$ and $n$ of the matrix;

- $test$: string representing the name of a test matrix from \cite{Zie}, or the empty string
(if the input matrix is not a test matrix);

- $sparse$: flag of type $tinytext$, possessing the value '0' if the matrix is non-sparse or one of the values, '1', '2', '3' for a sparse matrix;
this field is important for searching the matrices already stored in the database;

- $operation$: It defines operations on input matrices ($A+B$, $rA+sB$, $A^{-1}+B^{-1}...$) or defines the computation of
the weighted Moore-Penrose inverse;

- $matrix\_I$, $matrix\_II$ and $matrix\_III $: contain $IDs$ from the table $matrices\_in$ or default value '0'
in order to find a solution for the chosen operation and entered matrices;

- $r$, $s$, $p$ and $q$: optional coefficients used to determine matrix operations $rA$, $sB$, $A^{p}$ and $B^{q}$, respectively.

\smallskip
It is clear that dimensions of matrices and vectors used as arguments are limited by the maximal length of the type $longtext$, which is
within the interval $[0, 2^{32}-1]$;

\begin{exm}
In the present example $matrix\_A$ defines the test matrix $A$ from \emph {\cite{Zie}} in the particular case $a=1$,
$matrix\_B$ is randomly generated and the $matrix\_C$ is sparse and unstructured.

\begin{footnotesize}
$$matrix\_A=\left[
     \begin{array}{cccccccccc}
       11&10&9&8&7&6&5&4&3&2 \\
       10&10&9&8&7&6&5&4&3&2\\
9&9&9&8&7&6&5&4&3&2\\
8&8&8&8&7&6&5&4&3&2\\
7&7&7&7&7&6&5&4&3&2\\
6&6&6&6&6&6&5&4&3&2\\
5&5&5&5&5&5&5&4&3&2\\
4&4&4&4&4&4&4&4&3&2\\
3&3&3&3&3&3&3&3&2&1\\
2&2&2&2&2&2&2&2&1&0\\
1&1&1&1&1&1&1&1&0&-1\\
     \end{array}
   \right]_{11\times10}$$

$$matrix\_B=\left[
     \begin{array}{cccccccccc}
       282&-11&-206&-39&84&94\\
-11&241&-80&129&121&-86\\
-206&-80&306&4&-113&2\\
-39&129&4&394&-19&-219\\
84&121&-113&-19&119&15\\
94&-86&2&-219&15&184\\
     \end{array}
   \right]_{6\times 6}$$

$$matrix\_C=\left[
\begin{array}{cccccccccc}
1&2&3&0&0&0&0&0&0&0\\
0&2&0&0&0&0&0&0&0&0\\
1&0&0&4&0&0&0&0&0&0\\
0&0&0&0&0&0&0&0&0&0\\
0&4&0&0&0&8&0&0&0&0\\
0&0&0&0&0&0&0&0&0&0\\
0&0&0&0&0&0&0&0&0&0\\
0&0&0&0&0&0&0&0&0&0\\
0&0&0&0&0&0&0&0&0&0\\
0&0&0&0&0&0&0&0&0&2\\
     \end{array}
   \right]_{10\times10}$$
\end{footnotesize}

The corresponding fragment of the table {\it matrices\_in} is illustrated in Table 1.

\begin{center}\rm
\begin{footnotesize}
\begin{tabular}{|c|c|c|c|c|}\hline
\multicolumn{5}{|c|}{$ \bf matrices\_in$}\\ \hline \hline
$id\_in$ & $elements\_in$ & $dimension$ &   $test$ & $sparse$ \\\hline
$1$ & $11,10,9,..., 1,1,1,0,-1$ & $11\times 10$ &   $A\_10\_11$  & 0\\\hline
2 & 282,-11,-206,...,2,-219,15,184 & $6\times 6$ &   ' '  & 0\\\hline
3 & 0,0,0,1,2,2,4,4,9 & $10\times 10$   & ' ' & 1 \\\hline
4 & 0,1,2,1,0,3,1,5,9 & $10\times 10$   & ' ' & 2 \\\hline
5 & 1,2,3,2,1,4,4,8,2 & $10\times 10$   & ' ' & 3 \\\hline
\end{tabular}

Table 1. Matrices in the database table $matrices\_in$
\end{footnotesize} \end{center}

\noindent Three different matrices are stored in Table 1.
The $matrix\_A$, which represents the test matrix $A_{11\times 10}$ from \emph {\cite{Zie}},
is accommodated in the first row of the table.
The representation of the $matrix\_B$ is placed in its second row.
In the third, fourth and fifth rows is the sparse matrix representation for $matrix\_C$ in the $COO$ format.

\smallskip
The adjoint fragment of the table {\it matrices\_out} is illustrated by Table 2, where $A(-1)$ and $A(MN)$ denote the inverse and the weighted Moore-Penrose inverse, respectively.

\begin{center}\rm
\begin{footnotesize}
\begin{tabular}{|c|c|c|c|c|c|c|c|c|c|}\hline
\multicolumn{10}{|c|}{$ \bf matrices\_out$}\\ \hline \hline
$id\_out$\ & $elements\_out$ & $operation$ & $matrix\_I$ & $matrix\_II$ & $matrix\_III$   & $r$ & $s$ & $p$ & $q$ \\\hline
1 & 1,-1,0,...,-0.25,-0.417 & $A(-1)$ & 1 & 0 & 0  & 0 & 0 & 0 & 0  \\\hline
2 & 1974,-77,...,105,1288 & $r*A+s*B$ & 2 & 2 & 0  & 3 & 4 & 0 & 0 \\\hline
3 & 0,0,0,0,1,2,2,2,4,9 & $A*B$ & 3 & 3 & 0   & 0 & 0 & 0 & 0 \\\hline
4 & 0,1,2,3,1,0,1,2,1,9 & $A*B$ & 4 & 4 & 0   & 0 & 0 & 0 & 0 \\\hline
5 & 4,6,3,12,4,1,2,3,8,4 & $A*B$ & 5 & 5 & 0  & 0 & 0 & 0 & 0 \\\hline
4 & 0.755,-0.156,...,0.526,-0.2  & $A(MN)$ & 18 & 14 & 17 & 0 & 0 & 0 & 0 \\\hline
\end{tabular}

Table 2. Matrices in database table $matrices\_out$\end{footnotesize}\end{center}

\end{exm}

\subsection{Relations between matrix representations and matrix storage systems}

We present different methods for computing the pseudo-inverse $A^{\dagger}_{MN}$ and executing fundamental matrix operations, depending on the relationship between the implemented storage systems and matrix representations.
As we have already mentioned, matrices are stored in the database in three formats: the $R$ format, the $mR$ format and the $COO$ format (only for the sparse matrix).

\begin{figure}[ht]
  \begin{center}
    \includegraphics[width=6.8truecm]{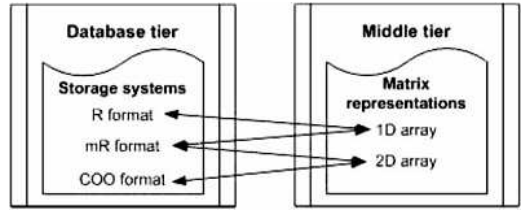}
    \caption{Relationship between storage systems and matrix representations} \label{FFig5}
  \end{center}
\end{figure}

To accelerate the search of the resulting matrices stored in the database
as well as to improve the pseudo-inverses computation, we exploit correlations between the presented storage systems indicated in Figure \ref{FFig5}.

\begin{figure}[ht]
  \begin{center}
    \includegraphics[width=12.7truecm]{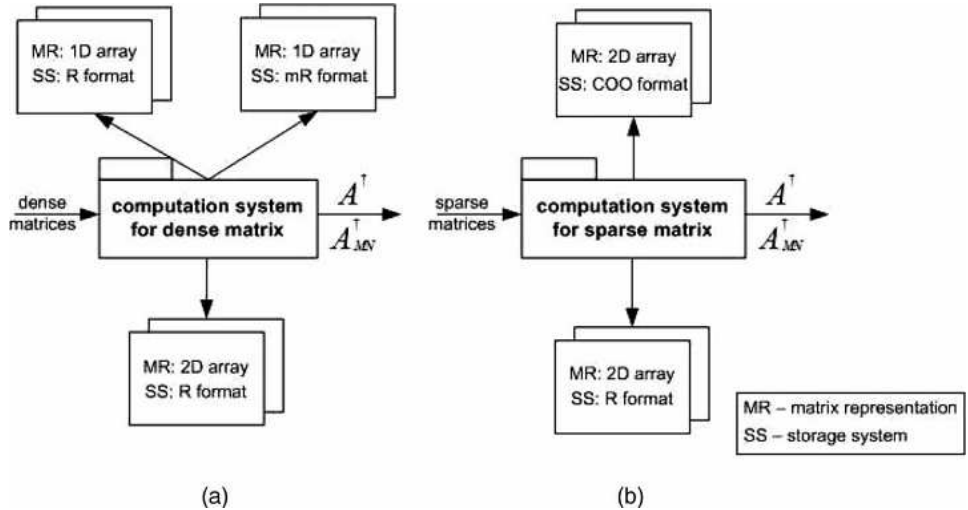}
    \caption{The computational systems for dense and sparse matrices} \label{FFig6}
  \end{center}
\end{figure}

We implemented different computational systems for dense as well as for sparse matrices in order to detect the best.
For dense matrices we use the computational systems illustrated on Figure \ref{FFig6}. a.
For sparse matrices we use a more suitable computational system, as shown on Figure \ref{FFig6}. b.

$UDFs$ for both dense and sparse matrices are implemented in the $2D$ array format.
$UDFs$ for sparse matrices are adapted to $COO$ format storing system, opposite to $UDFs$ for dense matrices, which are adapted to $R$ format.

\subsection{Implementation of matrix operations}

Implementation of an arbitrary matrix operation in the database tier can be described by the following algorithm.

\begin{algorithmbis} [Implementation of an arbitrary matrix operation.]
\begin{algorithmic}[1]
\REQUIRE Input matrices.
\STATE Define the input matrices and/or coefficients for processing and select the matrix operation.
\STATE Form the strings which define elements of entered matrices, applying Procedure \ref{p62} for each input (dense) matrix.
\STATE Search entered matrix in the database performing Procedure \ref{p63}
\STATE ${\bf if}$ the input matrices and required operations exist in the database {\bf then}
\STATE \      \ \ search the solution applying Procedure \ref{p65}
\STATE  \      \ \ ${\bf if}$ the solution exists  {\bf then}
\STATE  \      \ \  \ \ \ $\bf{goto}$ \emph{Step 16}
\STATE \      \ \    {\bf else}
\STATE  \      \ \  \ \ \ $\bf{goto}$ \emph{Step 14}
\STATE  \      \ \ ${\bf end}$ ${\bf if}$
\STATE {\bf else}
\STATE  \      \ \ Store entered matrix by applying Procedure \ref{p64} and \textbf{goto }\emph{Step 14}.
\STATE ${\bf end}$ ${\bf if}$
\STATE Processing the input matrix executing Procedure \ref{p67}.
\STATE Store the result performing Procedure \ref{p68}.
\STATE {\bf return} the result from the database using Procedure \ref{p66}
\end{algorithmic}\label{genalg}
\end{algorithmbis}

\noindent We provide some procedures, applicable to dense matrices, that are used in our implementation.

\smallskip
\begin{ppro} \label{p61} \rm Procedure \ref{p61} defines the input matrix by loading the corresponding $txt$ file.
It can be used in Step 1 of Algorithm \ref{genalg}.

\begin{footnotesize}
\begin{lstlisting}
function upload_file(){
   global $userfile,$userfile_name,$userfile_size,$userfile_type,$archive_dir,$WINDIR;
   if(isset($WINDIR)) $userfile=$str_replace("\\\\","\\",$userfile);
   $filename=basename($userfile_name);
   if($userfile_size <= 0) die("$filename is empty.");
   if(!@copy($userfile,"$archive_dir/$filename"))
   die("Can't copy $userfile_name to $filename.");
   if (!isset($WINDIR)&&!@unlink($userfile))
   die("Can't delete the file $userfile_name.");}
\end{lstlisting}
\end{footnotesize}
\end{ppro}

\begin{ppro} \label{p62} \rm  Code used in Step 2 of Algorithm \ref{genalg} for defining the string (vector) which includes entries of the dense matrix.

\begin{footnotesize}
\begin{lstlisting}
function make_vector($file){
   $m=count($file);//number of rows
   $string="";
   for($i=0;$i<$m;$i++){
       $file[$i]=rtrim($file[$i]).","; $string=$string.$file[$i];}
   $string1=$str_replace(" ",",",$string);
   $string2=substr($string1,0,strlen($string1)-1);//vector of all elements from matrix
   return $string2;}
\end{lstlisting}
\end{footnotesize}
\end{ppro}

\begin{ppro} \label{p63} \rm  The searching of a dense matrix with randomly chosen elements is implemented by the next function.
The search of a test matrix is defined by its unique name.

\begin{footnotesize}
\begin{lstlisting}
function if_exist($file){
   $db_link=db_connect();//connect on the database
   $m=m($file);//number of rows
   $n=n($file);//number of columns
   $dimension=$m."x".$n;  $string2=make_vector($file);
   $query="select * from MATRICES_IN where DIMENSION='$dimension'";
   $result=mysql_query($query,$db_link);  $sum=0;
   while($row=mysql_fetch_row($result)){
      if($row[1]==$string2){$sum++;  break;}}
   if($sum==1){
      $id_matrix=$row[0];}//matrix exist
   else
      $id_matrix=false;//matrix don't exist
   return $id_matrix;}
\end{lstlisting}
\end{footnotesize}
\end{ppro}

\begin{ppro} \label{p64} \rm  If the input matrix does not exist in the database, we store it applying the next code.

\begin{footnotesize}
\begin{lstlisting}
function enter_matrix($string2,$dimension,$test){
   $db_link=db_connect();
   $query="insert into MATRICES_IN(ELEMENTS_IN,DIMENSION,TEST)
   values('$string2','$dimension','$test')";
   $result=mysql_query($query,$db_link);}
\end{lstlisting}
\end{footnotesize}
\end{ppro}

\begin{ppro} \label{p65} \rm  If the entered matrices and operation already exist, we search the result from the database.

\begin{footnotesize}
\begin{lstlisting}
function result_exist($operation,$matrix1,$matrix2,$matrix3,$r,$s,$p,$q){
   $db_link = db_connect();
   $query = "select * from MATRICES_OUT where OPERATION ='$operation' and
   MATRIX_I='$matrix1' and MATRIX_II='$matrix2' and MATRIX_III='$matrix3'
   and R='$r' and S='$s' and P='$p' and Q='$q'";
   $result=mysql_query($query,$db_link);
   $row=mysql_fetch_array($result);
   if($row[1]==NULL){$matrix_out=false;}
   else{$matrix_out=explode(",",$row["ELEMENTS_OUT"]);}
   return $matrix_out;}
\end{lstlisting}
\end{footnotesize}
\end{ppro}

\begin{ppro} \label{p66} \rm  If the result exists in the database table, we display it.

\begin{footnotesize}
\begin{lstlisting}
function presented_result($result,$m,$n){
   for($i=0,$z=0;$i<$m;$i++){
      for($j=0;$j<$n;$j++,$z++){
         $result= round($result[$z],3);
         echo "<input name=elements[] type=text size=\"4\" maxlength=\"5\"
         value=\"$result\"  class=\"style21\"  readonly=\"readonly\"> ";}
      echo "<br>";} }
\end{lstlisting}
\end{footnotesize}
\end{ppro}

\begin{ppro} \label{p68} \rm  The next code is aimed to store the result of the matrix manipulation.

\begin{footnotesize}
\begin{lstlisting}
function enter_solution($result,$operation,$id_m1,$id_m2,$id_m3,$r,$s,$p,$q){
   $db_link=db_connect();
   $string=implode($result,",");
   $query="insert into
      MATRICES_OUT(ELEMENTS_OUT,OPERATION,MATRIX_I,MATRIX_II,MATRIX_III,R,S,P,Q)
   values('$string','$operation','$id_m1','$id_m2','$id_m3','$r','$s','$p','$q')";
   $result=mysql_query($query,$db_link);}
\end{lstlisting}
\end{footnotesize}
\end{ppro}

\subsection{The pseudo-inverse computation}

The $Php$ code which implements the weighted Moore-Penrose inverse $A_{M,N}^\dagger$ according to Algorithm \ref{algwang}.
Branching in the code is defined according to the selected storage system for our database table.

\begin{ppro} \label{p67} \rm  The main function for computation of the Weighted Moore-Penrose inverse.

\begin{footnotesize}
\begin{lstlisting}
function WeightedInverse($ArrayDataMatrixM,$ArrayDataMatrixA,$ArrayDataMatrixN){
    $rows=count($ArrayDataMatrixA); $columns=count($ArrayDataMatrixA[0]);
    $aa=ithCol(1,$ArrayDataMatrixA);
    if(isZero($aa)==0)
        $ar=Transpose($aa);
    else{
        $ta=Transpose($aa);
        $alb=MultiplicationMatrices($ta,$ArrayDataMatrixM);
        $ali=MultiplicationMatrices($alb,$aa);
        $inv=Inverse($ali);
        $ar=MultiplicationMatrices($inv,$alb);
    }
    for($i=2;$i<=$columns;$i++){
        $ii=ithCol($i,$ArrayDataMatrixA);
        $di=MultiplicationMatrices($ar,$ii);
        $fc=frstiCol($i-1,$ArrayDataMatrixA);
        $mm=MultiplicationMatrices($fc,$di);
        $ci=SubsMatrices($ii,$mm);
        $nim1=mMinus($ArrayDataMatrixN,$i);
        $nim1g=Inverse($nim1);
        $li=mColumn($ArrayDataMatrixN,$i);
        if(isZero($ci)==0){
            $nii=mskalar($ArrayDataMatrixN,$i);
            $dit=Transpose($di);
            $dtn=MultiplicationMatrices($dit,$nim1);
            $dtnd=MultiplicationMatrices($dtn,$di);
            $dtnd=matNum($dtnd);
            $ditl=MultiplicationMatrices($dit,$li);
            $lit=Transpose($li);
            $litdi=MultiplicationMatrices($lit,$di);
            $ditdi=SumaMatrices($ditl,$litdi);
            $ditdi=matNum($ditdi);
            $nim1li=MultiplicationMatrices($nim1g,$li);
            $litnli=MultiplicationMatrices($lit,$nim1li);
            $litnli=matNum($litnli);
            $lktar=MultiplicationMatrices($lit,$ar);
            $arnklk=MultiplicationMatrices($fc,$nim1li);
            $novo=MultiplicationMatrices($lktar,$arnklk);
            $novo=matNum($novo);
            $si=$nii+$dtnd-$ditdi-$litnli+$novo;
            $dtnar=MultiplicationMatrices($dtn,$ar);
            $dilit=SubsMatrices($dtnar,$lktar);
            $bit=ScalarMatrix($dilit,1/$si);
        }
        else{
            $cit=Transpose($ci);
            $citm=MultiplicationMatrices($cit,$ArrayDataMatrixM);
            $pom=MultiplicationMatrices($citm,$ci);  $pom=matNum($pom);
            $bit=ScalarMatrix($citm,1/$pom);
        }
        $nim1li=MultiplicationMatrices($nim1g,$li);
        $arai=MultiplicationMatrices($ar,$fc);
        $arnim1=MultiplicationMatrices($arai,$nim1li);
        $pi=SubsMatrices($nim1li,$arnim1);
        $k1=MultiplicationMatrices($di,$bit);
        $ark1=SubsMatrices($ar,$k1);
        $pib=MultiplicationMatrices($pi,$bit);
        $ar=SubsMatrices($ark1,$pib);
        $ar=appRow($ar,$bit);    }
    return $ar;}
\end{lstlisting}
\end{footnotesize}

\noindent Implementation of previous function requires several auxiliary procedures.

\smallskip
\noindent Function \emph{ithCol} generates the \(i\)th column \textit{a\(_{i}\)} of \(A\).

\begin{footnotesize}
\begin{lstlisting}
function ithCol($column,$ArrayDataMatrix1){
    $rows1=count($ArrayDataMatrix1); $columns1=count($ArrayDataMatrix1[0]);
    for($i=0;$i<$rows1;$i++) {
        for($j=0;$j<$columns1;$j++){
            if($j+1==$column)
            $ArrayCol[$i][0]=$ArrayDataMatrix1[$i][$j]; } }
    return $ArrayCol;}
\end{lstlisting}
\end{footnotesize}

\noindent The submatrix \(A_{j}\)=[\textit{a}\(_{1, ...,}\)\textit{a\(_{j}\)}] which contains first \(j\leq n\) columns of the matrix \(A=A_{n}\)=[\textit{a}\(_{1, ...,}\)\textit{a\(_{n}\)}] can be generated performing the next function:

\begin{footnotesize}
\begin{lstlisting}
function frstiCol($columns,$ArrayDataMatrix1){
    $rows1=count($ArrayDataMatrix1); $columns1=count($ArrayDataMatrix1[0]);
    for($i=0;$i<$rows1;$i++){
        for($j=0;$j<$columns1;$j++){
            if($j+1<=$columns)
            $ArrayCol[$i][$j]=$ArrayDataMatrix1[$i][$j];} }
    return $ArrayCol;}
\end{lstlisting}
\end{footnotesize}

\noindent Procedure \emph{matNum()} converts matrix of dimension $1\times 1$ into the number.

\begin{footnotesize}
\begin{lstlisting}
function matNum($ArrayDataMatrix1){
    $rows1=count($ArrayDataMatrix1); $columns1=count($ArrayDataMatrix1[0]);
    if($rows1==1&$columns1==1)
    return $ArrayDataMatrix1[0][0];}
\end{lstlisting}
\end{footnotesize}

\noindent Function \emph{isZero()} checks whether the parameter is zero matrix or not.

\begin{footnotesize}
\begin{lstlisting}
function isZero($ArrayDataMatrix1){
    $isZero=0; $rows1=count($ArrayDataMatrix1); $columns1=count($ArrayDataMatrix1[0]);
    for($i=0;$i<$rows1;$i++){
        for($j=0;$j<$columns1;$j++){
            if(round($ArrayDataMatrix1[$i][$j],3)!=0)
            $isZero++; }  }
    return $isZero;}
\end{lstlisting}
\end{footnotesize}

\noindent Function \emph{appRow()} appends the column vector $Y$ to the matrix $X$.

\begin{footnotesize}
\begin{lstlisting}
function appRow($ArrayDataMatrix1,$ArrayDataMatrix2){
    $isZero=0;
    $rows1=count($ArrayDataMatrix1);
    $columns2=count($ArrayDataMatrix2[0]);
    for($i=0;$i<$columns2;$i++){
        $ArrayDataMatrix1[$rows1][$i]=$ArrayDataMatrix2[0][$i];}
    return $ArrayDataMatrix1;}
\end{lstlisting}
\end{footnotesize}

\noindent Finally, the function \emph{Inverse()} returns the inverse of the input parameter.

\begin{footnotesize}
\begin{lstlisting}
function Inverse($ArrayDataMatrix){
    $rows=count($ArrayDataMatrix);  $columns=count($ArrayDataMatrix[0]);
    $nii=mskalar($ArrayDataMatrix,1);
    $N=1/$nii;  $N=array(array($N));
    for($i=2;$i<=$columns;$i++){
        $s=mskalar($ArrayDataMatrix,$i);
        $nii=array(array($s));
        $li=mColumn($ArrayDataMatrix,$i);
        $lit=Transpose($li);
        $giip=MultiplicationMatrices($lit,$N);
        $gii1=MultiplicationMatrices($giip,$li);
        $gii=SubsMatrices($nii,$gii1);
        $gii2=matNum($gii);  $gii=1/$gii2;
        $fip=MultiplicationMatrices($N,$li);
        $fi1=ScalarMatrix($fip,$gii);
        $fi=ScalarMatrix($fi1,-1);
        $fit=Transpose($fi);
        $E1=MultiplicationMatrices($fi,$fit);
        $E2=ScalarMatrix($E1,1/$gii);
        $E=SumaMatrices($N,$E2);
        $ArrayDataMatrix1=appRow($E,$fit);
        $giin=array(array($gii));
        $ArrayDataMatrix2=appRow($fi,$giin);
        $N=finaly($ArrayDataMatrix1,$ArrayDataMatrix2);    }
    return $N;}
\end{lstlisting}
\end{footnotesize}
\end{ppro}

\section{Examples}\setcounter{equation}{0}

\begin{exm}
In the next table, we compare the CPU time spent for searching the matrices stored in the database
by two different storage systems.

\begin{center}\rm
\begin{scriptsize}
\begin{tabular}{|c|c|c|}\hline
\multicolumn{3}{|c|}{\textbf{Search of stored matrices}}  \\ \hline
\textbf{number of matrices }& $R$ \textbf{format} & $mR$\ \textbf{format}  \\  \hline
$50$        & 0.102 sec.    & 0.647 sec. \\
$100$        &  0.166 sec.    & 0.660 sec. \\
$500$        &0.7909 sec.    &14.079 sec. \\
$1000$   &1.901 sec.     &28.650 sec. \\ \hline
\end{tabular}
\end{scriptsize}
\end{center}

\begin{center}
\begin{footnotesize} \rm Table 3. The CPU time for searching matrices dimension $70\times70$ in the $R$ and $mR$ format.\end{footnotesize}
\end{center}

\noindent This testing is performed for dense matrices with randomly chosen elements.
from Table 3. we conclude that the CPU time for searching is much shorter when the $R$ format is used.
\end{exm}

Let us mention that the searching of the test matrices from \cite{Zie} is based on the search by a unique name,
and does not depend on the number of the matrices stored in the database.

\begin{exm} In Table 4. are arranged $CPU$ times required for computation of the pseudo-inverse $A^{\dagger}_{MN}$.
Matrices are represented in two different ways: as $1D$ arrays based on the $mR$ and $R$ format,
or in the form of $2D$ arrays combined with the $R$ format storage system.

\begin{scriptsize}\rm
\begin{center}
\begin{tabular}{|c|c|c|c|}\hline
\multicolumn{4}{|c|}{$A^{\dagger}_{MN}$}  \\ \hline
\multicolumn{4}{|c|}{
\textbf{matrix representation: storage system}}  \\ \hline
\textbf{$m\times n$ }&
\textbf
\emph{\textbf{\emph{1D array}:}} \emph{$mR$ \textbf{format}} & \emph {\textbf{1D
array:}} \emph{$R$\ \textbf{format}} & \emph
{\textbf{2D array: R format}}  \\    \hline
$A\_30\_3$        &4.485 sec.     &2.440 sec.  &0.897 sec.\\
$A\_50\_4$        &32.834 sec.    &21.270 sec.  &6.199 sec.\\
$F\_15\_2$        &0.666 sec.     &0.342 sec.  &0.166 sec.\\
$F\_30\_3$        &4.379 sec.     &2.347 sec. &0.880 sec.\\
$F\_50\_4$        &32.564 sec.    &20.981 sec. &6.077 sec.\\
$S\_50\_4$        &32.597 sec.    &20.874 sec. &6.070 sec.\\
$S\_80\_5$        &269.282 sec.   &206.007 sec. &40.75 sec.\\
$50\times50$       &32.423 sec.    &21.294 sec. &6.203 sec.\\
$15\times15$       &0.632 sec.     &0.471 sec. &0.160 sec.\\
$50\times35$   &9.79 sec.      &5.587 sec. &1.997 sec.\\
$45\times70$   &125.18 sec.    &86.210 sec. &19.920 sec.\\
$60\times60$       &70.260 sec. &49.964 sec. &12.627 sec.\\
\hline
\end{tabular}
\end{center}
\end{scriptsize}
\begin{center}\rm \begin{footnotesize}Table 4. Computation time for $A^{\dagger}_{MN}$ when matrices don't exist in the database.\end{footnotesize}
\end{center}
Results of the testing show that the best processing time is achieved in the case when matrices are presented in the form of $2D$ arrays.
\end{exm}

\begin{exm}
In this example, we compare the CPU times obtained during the computation of the pseudo-inverse $A^{\dagger}_{MN}$
(in the case when the input matrices are not stored in the database) with the CPU times required for searching the same results (in the case when the input matrices are stored in the database).
Matrices are presented in $2D$ array representation, $R$ format storage system.

\begin{scriptsize}
\begin{center}\rm
\begin{tabular}{|c|c|c|}\hline
\multicolumn{3}{|c|}{
\textbf{A\(^{\dagger}_{MN}\)}}  \\ \hline
\multicolumn{3}{|c|}{
\textbf{matrix representation: storage system}}  \\ \hline
\textbf{$m\times n$ }&
\textbf{$2D$ $array: R format$ (exist)} & $2D$ $array:
R format$ (don't exist)
  \\
  \hline
$20\times20$   &  0.043 sec.  & 0.309 sec. \\
$30\times30$   &  0.051 sec.  & 2.105 sec. \\
$45\times45$   &  0.062 sec.  & 4.665 sec. \\
$50\times50$   &  0.070 sec.  & 6.729 sec.\\
$60\times60$   &  0.075 sec.  & 12.675 sec.\\
$70\times70$   &  0.079 sec.  & 26.614 sec.\\
$80\times80$   &  0.096 sec.  & 43.809 sec.\\\hline
\end{tabular}
\end{center}
\end{scriptsize}
\begin{center}\rm \begin{footnotesize}Table 5. The CPU time for $A^{\dagger}_{MN}$ when matrices exist and don't exist in the database.\end{footnotesize}
\end{center}

\noindent The CPU time required for searching the solution from the database is negligible with respect to the time needed for the recalculation, especially if the matrix has larger dimensions.
Let us mention that 10030 records are stored in the table $matrices\_in$ at the moment of testing.
\end{exm}

\begin{exm} Comparison of the $CPU$ time for executing fundamental matrix operations is given in Table 6.

\begin{scriptsize}\rm
\begin{center}
\begin{tabular}{|c|c|c|c|c|}\hline
\multicolumn{5}{|c|}{\textbf{Fundamental operation on matrices}}\\ \hline\hline
\multicolumn{2}{|c|}{\textbf{$m\times n$}} & \multicolumn{1}{|c|}{\textbf{$operation$}}&\multicolumn{1}{|c|}{\textbf{$1D$ $format$: $R format$}}&\emph{\textbf {2D
array: $R format$}}\\ \hline
$3 \times3$  & $3\times3$    & multiplication & 0.0468 sec.&
0.0043 sec.
\\
$5\times5$    & $5\times5$    & multiplication & 0.0481 sec.&
0.0042 sec.\\
$10\times10$    & $10\times10$    & multiplication & 0.0643 sec.& 0.0062 sec.\\
$20\times50$  & $50\times50$    & multiplication & 0.1531 sec.&
0.0778 sec.\\
$45\times45$   &  $45\times70$    & multiplication & 0.5618 sec. & 0.1369 sec.\\
$80\times80$    &$80\times60$    &multiplication & 0.9325 sec.&
0.7270 sec.\\
$80\times70$   &$70\times70$     &multiplication & 0.9675 sec.&
0.5600 sec.\\
$81\times81$   &  $81\times81$  &multiplication & 1.2831 sec.&
0.7470 sec.\\\hline
$3\times3$        & $3\times3$    & addition & 0.0425 sec.&
0.0040 sec.\\
$5\times5$        & $5\times5$    & addition & 0.0431 sec.&
0.0043 sec.\\
$10\times10$    & $10\times10$    & addition & 0.0431 sec.&
0.0048 sec.\\
$50\times50$    & $50\times50$    & addition & 0.1400 sec.&
0.0326 sec.\\
$60\times60$    &  $60\times60$    & addition & 0.2012 sec. & 0.0480 sec.\\
$70\times70$    &$70\times70$    &addition & 0.2481 sec.& 0.0705
sec.\\
$80\times80$   &$80\times80$     &addition & 0.3287 sec.& 0.0919
sec.\\\hline
$3\times3$        & $3\times3$    & substraction & 0.0475 sec.
& 0.0041 sec.\\
$5\times5$        & $5\times5$   & substraction & 0.0512 sec.&
0.0044 sec.\\
$10\times10$        & $10\times10$    & substraction & 0.0350 sec.& 0.0050 sec.\\
$50\times50$        & $50\times50$    & substraction & 0.1431 sec.& 0.0319 sec.\\
$60\times60$        &  $60\times60$    & substraction & 0.1875 sec. & 0.0733 sec.\\
$70\times70$        &$70\times70$    &substraction & 0.2512 sec.& 0.0415 sec.\\
$81\times81$   &$81\times81$     &substraction & 0.3150 sec.&
0.1196 sec.\\\hline
\end{tabular}
\end{center}
\end{scriptsize}
\begin{center}\rm \begin{footnotesize}Table 6. The CPU time for $1D$ array and $2D$ array in $R$ format processing.
\end{footnotesize}\end{center}

\noindent The arranged results show that the smallest computational time is obtained in the case when the matrix is presented
in the format of $2D$ array, $R$ format storage system.

\end{exm}
Testing was done on the local machine and from client in a wireless network. We had an access to the web server using the infrastructure mode wireless networking with an access point. Testing was executed on the server machine with: Windows edition: $Windows$ $Vista^{(TM)}$ $Ultimate$; Processor: $Intel(R)$  $Pentium(R)$ $Dual$ $CPU$ $T3200\ @\  2.00GHz$;
Memory $(RAM): 2940 MB$; System type: $32-bit\ Operating\ System$; Free Software: $WAMP$ $5$ $1.7.4$ installs: $PHP \ 5.2.5$, $Apache \ 2.2.6$ $Server$, $MySQL \ 5.0.45$ and $phpMyAdmin \ 2.11.2.1$.

\smallskip
The $CPU$ time shows that searching of a matrix which is given in the $R$ format
gives better results than the searching of the same matrix presented in the $mR$ format.
Furthermore, the best results are obtained in the case when matrices are represented in view of the two-dimensional arrays (on the middle tier), application logic adapted to the $R$ format, and matrices are stored in the $R$ format in the database.
Searching time is inversely proportional with the number of data transferred between the middle of the application tier and the database tier.

\section{Conclusions}

This research contributes to the development of an $OSS$ application and construction of the matrix
library, especially in computation of the generalized inverses.
Several storage technics are discussed and tested in the $PHP/MySQL$ environment.
Data's preparation as an important preprocessing step of data mining is a further application for
database matrix computations. There are several issues for future work.
We plan to develop mechanisms which decrease memory usage
and minimize the searching time in accordance with different storage systems.
Furthermore, it is possible to adapt $UDF$s to sparse structured matrices,
such as diagonal matrices, sparse symmetric matrices, triangular matrices, Toeplitz matrices, etc.
We need to identify other mathematical operations with wide applicability that can be implemented
with the $PHP/MySQL$ environment, thereby enhancing the $DBMS$ data mining functionality.
Also, further implementation of the matrix library will be based on the principles of $Object$-$Oriented$ $Programming$.

\end{document}